\documentclass[twocolumn,prb,aps,superscriptaddress,showpacs]{revtex4}

\usepackage{graphicx}
\usepackage{dcolumn}
\usepackage{amsmath}
\newlength{\figwidth}
\begin{document}


\title{Properties of charge density waves in La$_{2-x}$Ba$_{x}$CuO$_4$}
\author{Young-June Kim}
\email{yjkim@physics.utoronto.ca} \affiliation{Department of
Physics, University of Toronto, Toronto, Ontario M5S~1A7, Canada}
\author{G. D. Gu}
\affiliation{Department of Condensed Matter Physics and Materials
Science, Brookhaven National Laboratory, Upton, New York,
11973-5000}
\author{T. Gog}
\affiliation{CMC-CAT, Advanced Photon Source, Argonne National
Laboratory, Argonne, Illinois 60439}
\author{D. Casa}
\affiliation{CMC-CAT, Advanced Photon Source, Argonne National
Laboratory, Argonne, Illinois 60439}

\date{\today}

\begin{abstract}
We report a comprehensive x-ray scattering study of charge density
wave (stripe) ordering in $\rm La_{2-x}Ba_xCuO_4 (x \approx 1/8)$,
for which the superconducting $T_c$ is greatly suppressed. Strong
superlattice reflections corresponding to static ordering of charge
stripes were observed in this sample. The structural modulation at
the lowest temperature was deduced based on the intensity of over 70
unique superlattice positions surveyed. We found that the charge
order in this sample is described with one-dimensional charge
density waves, which have incommensurate wave-vectors (0.23, 0, 0.5)
and (0, 0.23, 0.5) respectively on neighboring $\rm CuO_2$ planes.
The structural modulation due to the charge density wave order is
simply sinusoidal, and no higher harmonics were observed. Just below
the structural transition temperature, short-range charge density
wave correlation appears, which develops into a large scale charge
ordering around 40 K, close to the spin density wave ordering
temperature. However, this charge ordering fails to grow into a true
long range order, and its correlation length saturates at $\sim
230\AA$, and slightly decreases below about 15 K, which may be due
to the onset of two-dimensional superconductivity.
\end{abstract}


\maketitle

\section{Introduction}

Over twenty years have passed since the discovery of
superconductivity in $\rm La_{2-x}Ba_xCuO_4$ (LBCO) by Bednorz and
Muller in 1986.\cite{Bednorz86} Although the theory of high
temperature superconductivity still remains elusive, much
understanding of this exciting condensed matter system, and also
physics of correlated electron systems in general, have been gained
in the past two decades. There have been many important experimental
discoveries and theoretical ideas to describe the observed physical
properties, and a number of recent review articles cover many of
these developments.\cite{Bonn06,Norman03,Lee06,Anderson04} One of
these is the so-called charge and spin stripe picture. That is,
doped holes in the cuprates tend to aggregate in one-dimensional
domain walls separating regions of antiferromagnetically ordered
spin domains. As early as 1988, incommensurate spin fluctuations in
cuprate superconductors were observed using neutron
scattering.\cite{Yoshizawa88,Thurston89,Cheong91} These observations
stimulated various theoretical work dealing with charged domain
walls in two dimensional Hubbard
model.\cite{Schulz89,Poilblanc89,Zaanen89} In 1995, motivated by the
theoretical model put forward by Emery and
Kivelson,\cite{Emery93,Kivelson94} Tranquada and coworkers
discovered static ordering of charge and spins in $\rm
La_{1.6}Nd_{0.4}Sr_{0.125}CuO_4$ (LNSCO) that is consistent with the
stripe model.\cite{Tranquada95} The next breakthrough was the
discovery of static ordering of spin stripes in $\rm
La_{2-x}Sr_{x}CuO_{4+y}$ (LSCO) over a wide doping
range.\cite{Kimura98,Wakimoto99,Lee99} In addition to these static
incommensurate ordering of spins in LSCO, incommensurate spin
fluctuations have been observed in a number of materials, including
$\rm YBa_2Cu_3O_{6+x}$ (YBCO).\cite{Dai98}

Despite such extensive experimental work on the incommensurate spin
fluctuations and ordering in cuprate superconductors, experimental
studies on the charge counterpart have been relatively scarce. In
recent years, the scanning tunneling spectroscopy (STS) technique
has attracted much attention due to its ability to provide real
space image of charge distribution on $\rm
Bi_2Sr_2CaCu_2O_{8+\delta}$ and $\rm (Ca,Na)_2CuO_2Cl_2$, etc.
Although these STS studies provide unprecedented information on the
inhomogeneous distribution of charge density and superconducting
gap,\cite{Pan01,Hoffman02a,Hoffman02b,Hanaguri04} due to the surface
sensitive nature of the technique, its application has been so far
limited to a subset of cuprate samples. In contrast, neutron
scattering investigation of incommensurate correlation has been
observed in many different materials. Although there have been
recent neutron scattering studies on YBCO, the bulk of studies on
stripes have been carried out with LSCO and
LNSCO.\cite{Yamada98,Kimura98,Wakimoto99,Tranquada95} Recently, it
has become possible to synthesize high quality large single crystals
of LBCO, and various experimental studies have been carried out to
elucidate the charge and spin properties of
LBCO.\cite{Tranquada04,Abbamonte05,Reznik06,Homes06,Valla06,Li07,LBCO-field}

However, neutron scattering investigation of the charge counterpart
of the stripe order has been very limited, due to its indirect
coupling to charge degrees of freedom, and also its low signal to
noise ratio. On the other hand, x-ray scattering couples directly to
the charge degree of freedom, and is still a bulk probe (5-10
microns). Except for when the incident photon energy is near the
absorption edges, the largest contribution to the x-ray scattering
intensity comes from the structural modulations accompanying charge
order. In this sense, x-ray scattering is similar to neutron
scattering, but one can use a synchrotron source and obtain very
high intensity and high momentum resolution. The first x-ray study
of the charge stripes was done with very high energy x-rays
(E=100keV) at HASYLAB by Zimmermann and
coworkers.\cite{Zimmermann98} They were able to elucidate the
stacking structure and temperature dependence of the charge stripe
ordering in their study of LNSCO. Further studies of LNSCO x=0.15
sample have been carried out at the same
facility.\cite{Niemoller99,Wakimoto03} In addition, the structural
phase transition and its relationship with the charge ordering has
been investigated with regular (low energy) x-ray scattering on $\rm
La_{1.875}Ba_{0.125-x}Sr_{x}CuO_4$ (LBSCO).\cite{Kimura03} Recent
availability of the LBCO crystals have made it possible to carry out
more detailed investigation of LBCO using soft x-ray resonant
scattering.\cite{Abbamonte05}

In this paper, we report our comprehensive investigation of charge
stripe ordering in LBCO using synchrotron x-ray scattering. As
previously reported, clear peaks corresponding to the charge stripe
ordering were observed. We confirmed some of the earlier
observations for LNSCO in the current LBCO sample, which indicates
that the charge stripe ordering in these two compounds are very
similar. In addition, we were able to study the structure of the
charge ordering in detail by surveying the intensity of various
charge stripe superlattice peaks. We find that the charge stripe
order is described with a sinusoidal charge density wave with
incommensurate wave-vector (0.23, 0, 0.5) in this sample, and no
higher harmonics were observed. Short-range charge stripe
correlation sets in as soon as the structural transition occurs in
this sample and the stripe correlation grows into a large scale
charge ordering around 40 K, which is close to the spin density wave
ordering temperature. However, this charge ordering fails to grow
into a true long range order, and its correlation length saturates
at 230\AA, and slightly decreases below about 15 K, which may be due
to the onset of two-dimensional (2D) superconductivity as suggested
by Li et al.\cite{Li07}

After first describing the experimental details in the next section,
we will present our results on the structural phase transition,
incommensurability, charge order structure, and temperature
dependence in order. Some of the implications of these experimental
results will be discussed in Section VI.

\section{Experimental details}

Preliminary x-ray scattering experiments were carried out at the
X22C beamline at National Synchrotron Light Source (NSLS).
Additional high-resolution data were obtained at the 9ID (CMC-CAT)
beamline at Advanced Photon Source (APS). In both experiments, the
incoming photon was monochromatized by a double bounce Si(111)
monochromator in vertical scattering geometry. Either graphite (002)
or Si (111) reflections were used to reduce the background and also
to improve the momentum resolution. Data were obtained both in the
(HK0) and the (H0L) scattering plane with either 8.9 keV (NSLS) or
12 keV (APS) photons. The samples used in our measurements were
grown with traveling-solvent floating zone technique. A small piece
from a large boul of single crystals was cut to reveal the [0KL]
surface, which is polished. In the temperature range studied, the
LBCO sample is in its low-temperature tetragonal (LTT) phase.
However, for consistency, we will use the notation following the
high-temperature tetragonal (HTT) lattice parameters, with $a=3.787$
\AA\ and $c=13.24$ \AA.

The superconducting transition temperature of the sample was
determined by measuring diamagnetic signal with Quantum Design MPMS
SQUID magnetometer, as shown in the Fig.~\ref{fig1} inset. The onset
temperature is about $\sim 6$ K. We have not carried out further
characterization to determine the doping level accurately, but from
the suppression of $T_c$, we can estimate that the doping is not
very different from x=0.125. All our x-ray measurements were carried
out on a single piece of crystal, and the magnetization was measured
with a small piece cut from this original piece.

\section{Low temperature tetragonal structure}
\label{sec:ltt}

One of the most important aspects of LBCO (and also LNSCO) that
distinguishes this from other cuprate superconductors is its low
temperature structure. Specifically, LBCO has the so-called LTT
structure, which can stabilize static ordering of charge stripes. It
is well-known that the crystal structure of LSCO at high temperature
is tetragonal with flat $\rm CuO_2$ planes (Symmetry $I4/mmm$). As
temperature is lowered, structural phase transition to an
orthorhombic phase called low-temperature orthorhombic (LTO,
symmetry group $Bmab$) occurs, in which the $\rm CuO_2$ plane is
buckled, such that the Cu-O-Cu bond angle deviates from 180
degrees.\cite{Birgeneau87} One should note that the buckling is
caused by rotation of $\rm CuO_6$ octahedra along the axis making 45
degrees with the Cu-O-Cu direction. (See Fig.~2(a)) This is the low
temperature structure of doped LSCO samples as well, at least up to
x=0.2. For $x > 0.2$, the low temperature structure remains
tetragonal. LBCO also goes through the same structural phase
transition from high temperature tetragonal (HTT) phase to low
temperature orthorhombic (LTO) phase at $\sim 200$ K.
\cite{Axe89,Katano93}

\begin{figure}
\begin{center}
\includegraphics[width=\figwidth]{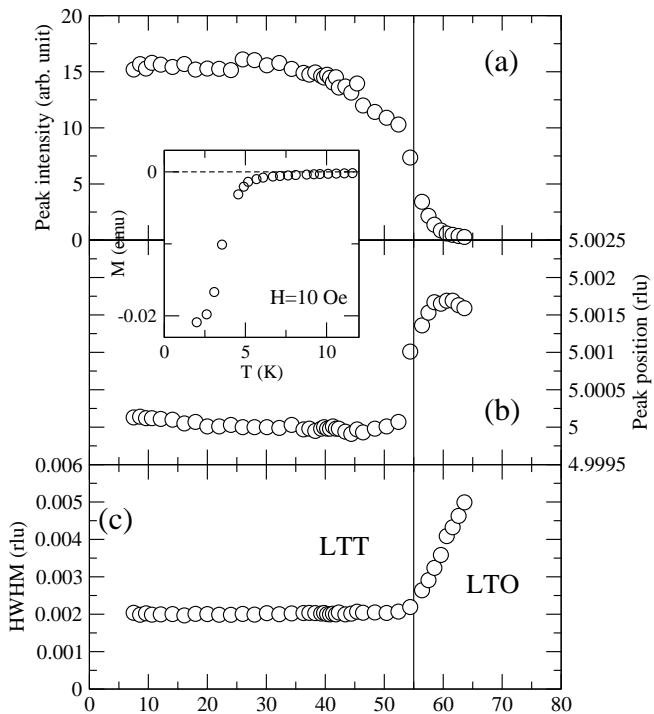}
\end{center}
\caption{The temperature dependence of the (a) peak intensity, (b)
nominal peak position, and (c) peak width of the {(5 0 0)} LTT
superstructure peak. The vertical line corresponds to the LTT-LTO
structural phase transition temperature $T_s$. Inset: Temperature
dependence of magnetization showing superconducting transition.}
\label{fig1}
\end{figure}
\begin{figure}
\begin{center}
\includegraphics[width=\figwidth]{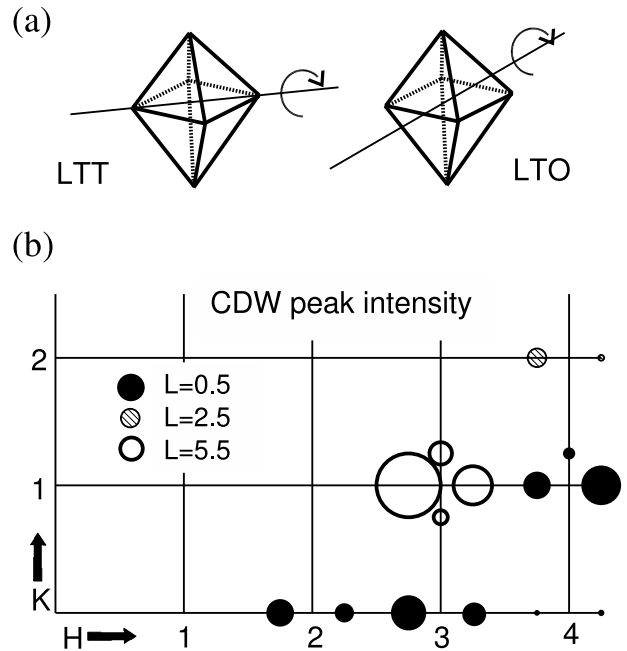}
\end{center}
\caption{(a) Schematic representation of the rotation of $\rm CuO_6$
octahedra in the LTT phase (left) and the LTO phase. (b) Reciprocal
space diagram to show the intensity distribution of the charge order
peaks. The location and the radii of the circle correspond to the
position and intensity of the observed peak. Note that satellite
peaks along the $K$ direction are only observed around the Bragg
peaks with non-zero $K$ values, as described in the text.}
\label{fig2}
\end{figure}

In addition to this HTT-LTO transition, LBCO goes through another
structural phase transition around 55K, and the low temperature
phase of LBCO again has tetragonal structure. However, this LTT
structure (symmetry group $P4_2/ncm$) is different from the HTT
phase, and the $\rm CuO_2$ plane remains buckled. The tetragonal
symmetry (with larger unit cell) is achieved by additional tilting
of the $\rm CuO_6$ octahedra. This time, the rotation axis is along
the Cu-O-Cu bond direction [See Fig. 2(a)]. As a result of this
rotation, stripes running along the Cu-O-Cu bond direction can be
pinned to the lattice. This is believed to be the origin of static
ordering of charge and spin stripes in LBCO and LNSCO. In
Fig.~\ref{fig1}, the temperature dependence of the (5 0 0)
superstructure peak is presented. The peak intensity drops abruptly
as temperature is raised above $T_s=55$ K, while the nominal peak
position suddenly jumps to higher value at the same temperature,
both signaling discontinuous structural phase transition. However,
the peak width is resolution limited at $T < T_s$, and grows
continuously for $T > T_s$. Note that this peak is resolution
limited in all directions at low temperatures even with the best
resolution setup, indicating that the LTT phase is established over
1400\AA.

Earlier neutron and electron diffraction studies on poly-crystalline
samples have reported that the LTT phase is short-range ordered, and
the LTO phase persists even at the lowest temperature
studied,\cite{Axe89,Zhu94} while the neutron diffraction study by
Katano and coworkers found that LTT fraction is much larger than LTO
fraction.\cite{Katano93} Our results seem to be consistent with the
latter result. That is, the LTT phase is developed fully at low
temperatures, over a macroscopic region. The temperature dependence
of the correlation length suggests that the LTT phase nucleates at
temperatures above $T_s$, and grows in size until it reaches
macroscopic scale at temperatures between 50 K and 60 K. The salient
point here is that the LTT phase is fully developed by the time the
temperature reaches the charge ordering temperature, and that the
LTT order is long-ranged, while the charge order is short-ranged as
we will show later.

\begin{figure}
\begin{center}
\includegraphics[width=\figwidth]{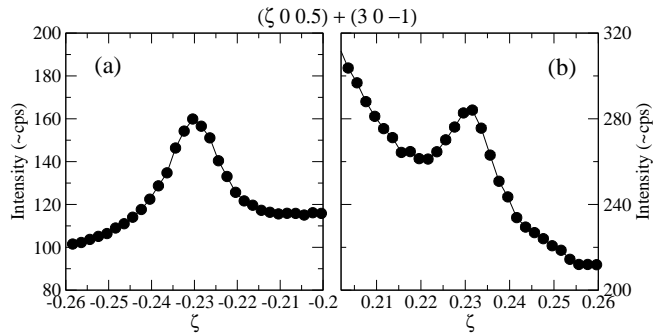}
\end{center}
\caption{Representative scans through the CDW peaks near the ${\bf
G}=(3, 0, \bar{1})$. Scans along the H direction on either side of
the Bragg peak are shown.} \label{fig3}
\end{figure}

\begin{table*}
\caption {Summary of the experimentally observed CDW peaks at
$\bf{G}+(\epsilon, 0, 0.5)$ and SDW peaks at ($0.5+\delta,
0.5+\delta,0)$ for a number of cuprates.} \label{table1}
\begin{ruledtabular} \begin{tabular}{clllcc}
Sample & doping (x) & $\epsilon$ CDW & $\delta$ SDW &
Note & Reference \\
\hline
LNSCO & 0.12 & 0.236 &  0.118 & neutron & \onlinecite{Tranquada95}\\
LNSCO & 0.12 & 0.236 &   & x-ray & \onlinecite{Zimmermann98}\\
LNSCO & 0.15 & 0.256 &   & x-ray & \onlinecite{Niemoller99}\\
LNSCO & 0.15 & 0.258 &  0.134 & x-ray+neutron & \onlinecite{Wakimoto03}\\
LBCO & 0.125 & 0.236 &  0.118 & neutron & \onlinecite{Fujita04}\\
LBCO & 0.125 & 0.245 &   & rsxs & \onlinecite{Abbamonte05}\\
LBSCO & 0.125 & 0.24 &  & x-ray & \onlinecite{Kimura03}\\
LBCO & 0.12 & 0.230 &  & x-ray & This work\\
\end{tabular}
\end{ruledtabular}
\end{table*}

\section{Ground state}

We were able to survey about 70 unique reciprocal lattice positions
where charge order peak is expected to exist. These are $\bf{G} \pm
{\bf g}_1$ and $\bf{G} \pm {\bf g}_2$, where $\bf{G}$ is a Bragg
peak position, and ${\bf g}_1=(\epsilon,0,0.5)$ and ${\bf
g}_2=(0,\epsilon,0.5)$ are incommensurate charge order wavevectors.
Some of these peak intensities are reported in Fig.~\ref{fig2}(b),
in which the size of a circle is proportional to the peak intensity
at this position. These measurements were carried out around
$T=10$~K, which is above the superconducting transition temperature.
In this section we present this survey result along with its
implication for the ground state structure of the charge stripe
order in LBCO.

\subsection{Incommensurate charge density wave}

In Fig.~\ref{fig3}, we show representative scans through the charge
order peak obtained around ${\bf G}=(3, 0, \bar{1})$. Scans along
the longitudinal direction (H) on either side of the Bragg peak are
shown in Fig.~\ref{fig3}(a) and (b), respectively. It is clear that
the charge order peak is located at the incommensurate wavevector,
${\bf g}_1=(\epsilon,0,0.5)$, with $\epsilon=0.230(2)$. That is, the
periodicity of the charge order is not exactly 4 times the lattice
constant, and it is indeed incommensurate with the lattice.
Therefore, the ``charge stripe order" in LBCO is better described as
an incommensurate charge density wave (CDW). The value of $\epsilon$
is consistently around $\sim 0.230$ for all the peaks studied in our
experiments. This incommensurability shows very little temperature
dependence over the whole range studied in our experiments. This is
also consistent with the neutron scattering results of Fujita et
al., in which they observed $\epsilon=0.236$ for the LBCO sample
with $T_c \approx 4$ K.

In the conventional stripe picture for x=1/8, one can construct a
charge order model with exactly $4a$ periodicity due to the
commensurate nature of the doping. That is, if each stripe carries
charge of half an electron as shown in recent x-ray
study,\cite{Abbamonte05} then the charge order unit cell is enlarged
by fourfold, and CDW peak at the incommensurability of
$\epsilon=0.25$ should be observed. In Table~\ref{table1},
experimental values for the CDW incommensurability in a number of
samples are summarized. Although the CDW and SDW incommensurability
are clearly correlated ($\epsilon=2\delta$), no obvious correlation
between the doping (x) and $\epsilon$ is found, and none of the data
shows exactly $\epsilon=0.25$. Empirically, $\epsilon < 2x$ is
observed in all the studies. Our observation of $\epsilon=0.230$
therefore is consistent with previous measurements. One explanation
for this empirical observation may come from doping dependence of
the incommensurability. It is well known that SDW incommensurability
in LSCO samples follow $\delta=x$ in the underdoped regime,
\cite{Yamada98} and if one assumes that $\epsilon=2x$ should hold in
the CDW order, our sample may have $x=0.115$ rather than nominal
$x=0.125$. The superconducting $T_c$ of LBCO samples are rapidly
suppressed near the $x=1/8$ doping, and $T_c$ can be a good measure
of actual doping level of the sample. From the phase diagram of
LBCO,\cite{Axe89} the $x=0.115$ sample should have $T_c$ between 5
and 10K, which is consistent with the measured value.

\begin{figure}
\begin{center}
\includegraphics[width=\figwidth]{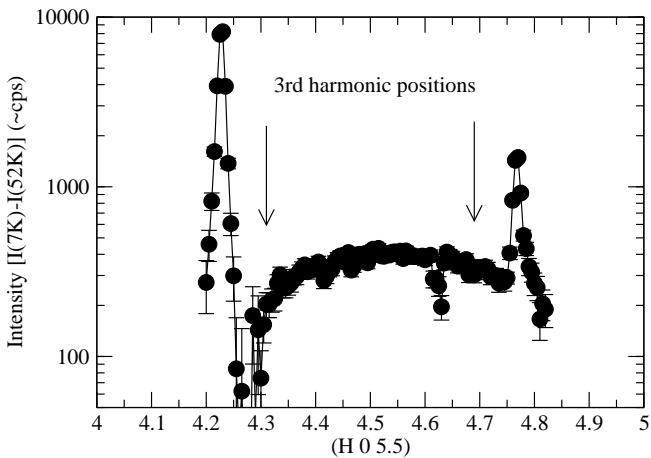}
\end{center}
\caption{A scan along the (H 0 5.5) direction. In order to show the
intensity due to the CDW order, intensity difference between the
same scan obtained at two different temperatures (7 K and 52 K) is
plotted. Two strong peaks correspond to the main CDW peaks at $(\pm
\epsilon, 0, 0.5)$. Logarithmic scale is used to emphasize the lack
of any higher harmonic peaks. The arrows denote expected positions
for third harmonic peaks: $(5,0,7)-3{\bf g}_1$ and $(4,0,4)+3{\bf
g}_1$.} \label{fig4}
\end{figure}

In Fig.~\ref{fig4}, we show the Q dependence of the scattering
intensity at (H 0 5.5) between H=4.2 and 4.8. In order to display
the scattering intensity due to the CDW at $T=10$ K, we have
subtracted the temperature independent background intensity at
$T=52$ K. Note that the vertical axis is shown in logarithmic scale
to magnify any small intensity. Considering the background noise, we
should be able to detect any higher harmonic peaks if they were
larger than 1 \% of the first harmonic. This observation strongly
suggests that the structural modulation due to CDW can be well
described by a simple sinusoidal modulation with doping-dependent
incommensurate wavevector (0.23, 0, 0.5).

\subsection{Structural modulation}

In Fig.~\ref{fig2}(b), we show intensity distribution of observed
superlattice peaks in the two-dimensional plane of (H K) for a
number of L values. One should notice that while the CDW peaks of
$(H, 0, 0) \pm {\bf g}_1$ are observed, $(H, 0, 0) \pm {\bf g}_2$
CDW peaks are not observed. That is, only a pair of superlattice
peaks are observed around the $(H, 0, 0)$ Bragg peak. When there is
a finite $K$ component, small ${\bf g}_2$ type CDW peaks could be
observed: e.g., $(3, 1 \pm \epsilon, 5.5)$ and $(4, 1+\epsilon,
2.5)$. The superlattice peak intensity arising from a structural
modulation vector, ${\bf u}_r$, is generally proportional to $({\bf
Q \cdot u}_r)^2$. Since ${\bf Q}$ is along the ${\bf a}^\ast$
direction for ${\bf G}=(H, 0, 0)$, ${\bf g}_2$-type superlattice
peaks are suppressed. Robertson and coworkers recently proposed that
one of the experimental signatures of 1D stripe charge ordering is 4
superlattice peaks, while 8 superlattice peaks are expected for a
checkerboard type charge order.\cite{Robertson06} We have also
searched for additional superlattice peaks along the diagonal
direction ${\bf q}=(\pm \eta, \pm \eta, l)$ around ${\bf G}=(4, 2,
6)$ at low temperatures without success. Therefore, our result seems
to indicate 1D nature of stripes.

The stacking structure of these 1D stripes was proposed by Tranquada
and coworkers, and confirmed in the x-ray study by Zimmermann and
coworkers. Specifically, if the $\rm CuO_6$ octahedra in one copper
oxygen layer (z=0) is tilted along the a-direction, the octahedra in
the neighboring layer (z=0.5) is tilted along the b-direction. This
naturally stabilizes the stripes running along the b-direction and
the a-direction for z=0 and z=0.5, respectively. Therefore, normally
one would expect to observe four satellite reflections, were it not
for the $({\bf Q \cdot u}_r)^2$ factor. The Coulomb interaction
between charge stripes causes the stripes in the z=1 layer to be
located in between the z=0 stripes, resulting in 2 unit cell (4
layer) periodicity along the c-direction. As a result, all the
stripe peaks are observed at $L=0.5, 1.5, ...$.

\begin{figure}
\begin{center}
\includegraphics[width=\figwidth]{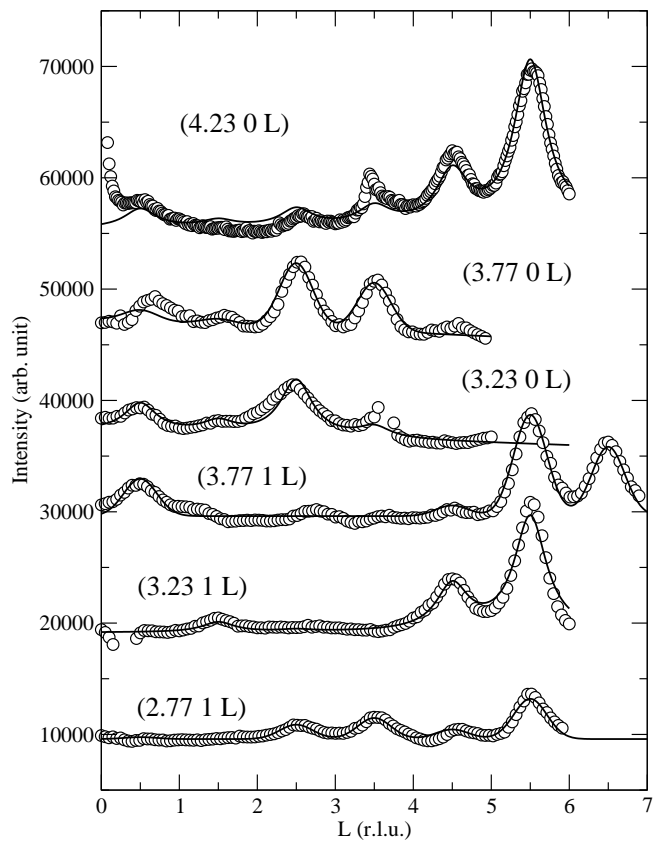}
\end{center}
\caption{L-dependence of the CDW peaks. The in-plane positions are
labeled. The difference between the scans taken at 7 K and 55 K are
plotted, and each scan is shifted for clarity.} \label{fig5}
\end{figure}

The intensity modulation along the L-direction is non-monotonic as
shown in Fig.~\ref{fig5} for a number of fixed in-plane wavevectors.
This observation is somewhat puzzling, since one expects the above
mentioned $({\bf Q \cdot u}_r)^2$ dependence. The main reason for
the non-trivial L-dependence of the intensity modulation is the
position of La and apical oxygens due to the tilting of octahedra in
the LTT phase. Specifically, in LTT phase, the La atoms are slightly
displaced from its original position in the LTO phase, so that they
no longer are located on top of the Cu atom. Instead, the La atoms
are found at $(\pm \delta, 0, \pm z_{La})$ and $(\frac{1}{2},
\frac{1}{2} \mp \delta, \frac{1}{2} \mp z_{La})$, with
$\delta=0.0106$ and $z_{La}=0.36$.\cite{Katano93} As a result the
structure factor of the CDW peaks acquire $\Delta z_{La} \sin
2\pi(z_{La}L - \delta h)$ and $ \Delta z_{La} \sin 2\pi(z_{La}L -
\delta k)$ factors, where $\Delta z_{La}$ is the c-axis modulation
amplitude of the La atoms.

In their neutron scattering study of charge ordering in $\rm La_2
NiO_{4.125}$, Tranquada et al. used a simple incommensurate
modulation model to fit the observed superlattice peak intensities,
and elucidate atomic displacements in the modulated
structure.\cite{Tranquada95b} However, unlike neutron scattering, it
is notoriously difficult to obtain absolute x-ray scattering
intensity over a wide $\bf{Q}$ range, due to the sample absorption
and the $\bf{Q}$-dependence of the atomic form factor. As a result,
although we followed the analysis in Ref.~\onlinecite{Tranquada95b}
to model the structural modulation due to the charge stripe order in
LBCO, we left the overall amplitude as a fitting parameter.
Refinement of the structural modulation model is beyond the scope of
this work, and will be addressed in a future publication.

We focus here on reproducing the L-dependence of the charge order
peak intensity. Following the treatment in
Ref.~\onlinecite{Tranquada95b}, we carried out similar least squares
fitting analysis of the observed peak intensities (shown in Fig.
2(b) and Fig. 5). It turns out that the fitting result is rather
insensitive to the modulation of oxygen atoms, as expected from
their small x-ray form factor. We also noticed that the in-plane
motion of La atoms is negligible, and the most important modulation
is the in-plane modulation of copper and out-of-plane (c-direction)
modulation of the La atoms. The solid lines shown in Fig.~\ref{fig5}
are produced with the least squares fitting results for the
intensity of observed CDW peaks. The background and the overall
intensity were varied to match the observed data, and the widths of
the peaks are fixed at 0.25 r.l.u. As one can see in Fig.~\ref{fig5}
that the L-dependent intensity modulation is well described by our
analysis.

\begin{figure}
\begin{center}
\includegraphics[width=\figwidth]{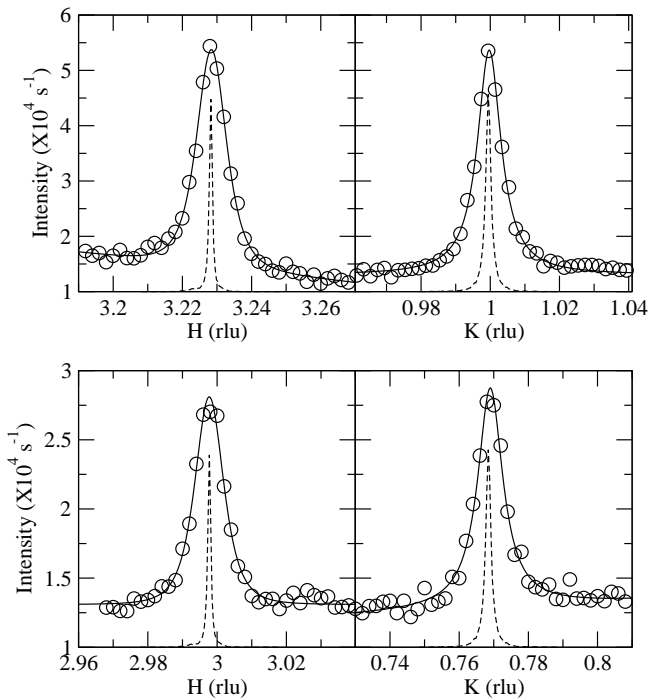}
\end{center}
\caption{Representative H and K scans around $\bf{G}=(3,1,5)$
obtained with high-resolution setup at 7 K. The solid lines are fits
to 2D Lorentzian as described in the text, and the dashed lines are
instrumental resolution.} \label{fig6}
\end{figure}

\subsection{Correlation length}

In Fig.~\ref{fig6}, we show representative scans at two of the
positions marked in Fig.~\ref{fig2}(b). The solid line is a fit to
2D Lorentzian line shape. One can clearly see that the peak width is
much broader than the instrumental resolution (dashed lines) as
measured at the nearby Bragg peak position. In addition, one can see
that the CDW peaks are quite isotropic in both H and K direction. In
fact, our fitting results give consistently similar widths along the
H and K direction in almost all measured peaks. The correlation
length, which is obtained as an inverse of the the Lorentzian peak
width is about 200-250 \AA. This in-plane isotropy of the CDW peak
may seem puzzling, considering the 1D nature of the CDW order, as
shown in previous section. However, this isotropic correlation
length can be understood if one considers the alternating stripe
directions in neighboring layers. In other words, the stripe ordered
region always has both H-stripe and K-stripe component, which makes
it isotropic.

We note that the observed correlation length is of similar order of
that observed by Abbamonte et al.,\cite{Abbamonte05} but much longer
than those in LNSCO,\cite{Tranquada99} presumably reflecting better
crystalline quality of the LBCO sample, compared to LNSCO. This is
also consistent with the neutron scattering measurements by Fujita
and coworkers,\cite{Fujita04} although the momentum resolution in
their study was quite poor. We emphasize however that the charge
stripe correlation length is not limited by the structural disorder.
As shown in Sec.~\ref{sec:ltt}, the LTT-LTO structural phase
transition is of first order nature, and the LTT superlattice peak
is resolution limited at all temperatures.

\section{Temperature dependence}

\subsection{CDW phase transition}

\begin{figure}
\begin{center}
\includegraphics[width=\figwidth]{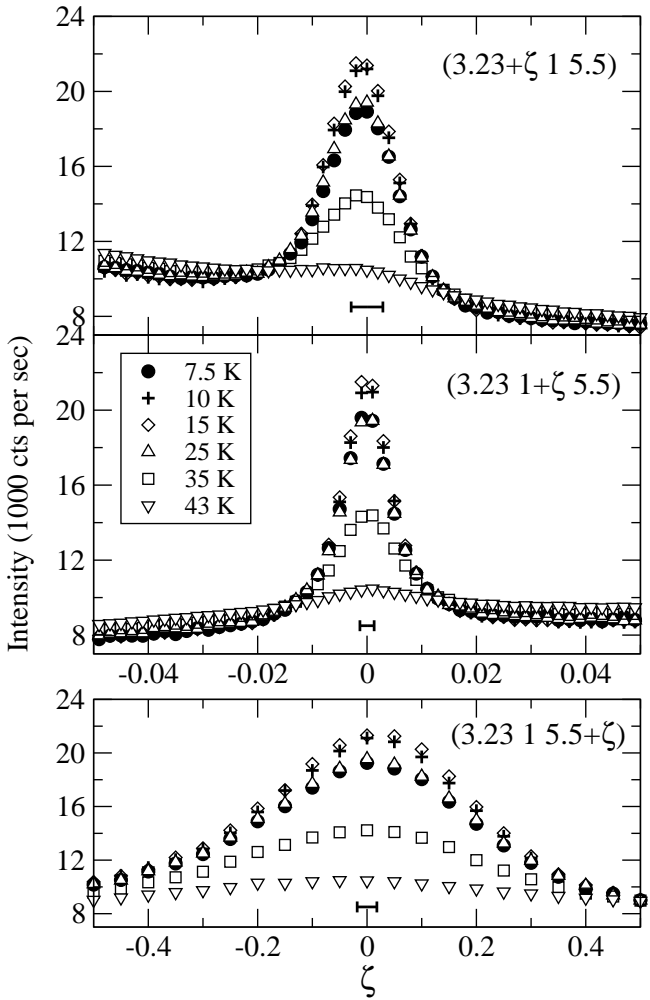}
\end{center}
\caption{Scans across the (3.23, 1, 5.5) CDW peak at different
temperatures. From top to bottom, H, K, and L scans are plotted as a
function of deviation, $\zeta$, from the peak position. The
horizontal bars denote the instrumental resolution. } \label{fig7}
\end{figure}

\begin{figure}
\begin{center}
\includegraphics[width=\figwidth]{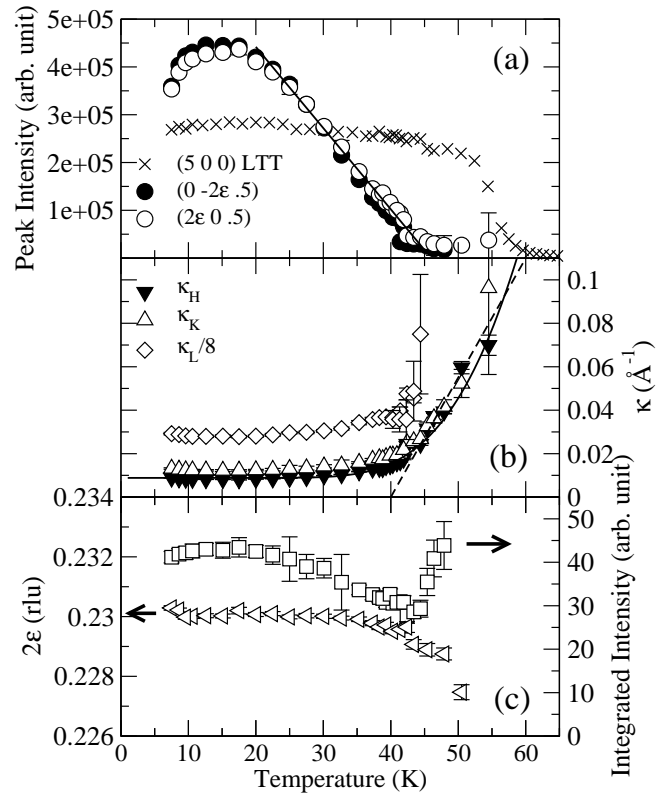}
\end{center}
\caption{Temperature dependence of the fitting parameters for the
CDW peak intensities around the (3, 1, 5) position are shown. (a)
The open and closed symbols are the peak intensities for the (3.23,
1, 5.5) and the (3, 0.77, 5.5) peak, respectively. Also shown in
this figure is the temperature dependence of the LTT superlattice
peak (5, 0, 0). (b) The peak widths (inverse correlation length)
along the H and K directions are shown in closed and open triangles,
respectively. These are obtained by scanning the (3.23, 1, 5.5) peak
along the H and K directions. The other satellite peak widths
exhibit identical temperature dependence. The peak widths along the
L-direction are shown in diamond. The solid and dashed lines are
fits described in the text. (c) The temperature dependence of the
incommensurability $2 \epsilon$ is plotted as triangles, while that
of the integrated intensity is plotted with squares. Both quantities
were obtained at the (3.23, 1, 5.5) position.} \label{fig8}
\end{figure}

In Fig.~\ref{fig7}, we show the temperature dependence of the
$(3.23, 1, 5.5)$ CDW superlattice peak. From top to bottom, H, K,
and L scans obtained at different temperatures are shown with
different symbols. The horizontal bars in each panel denote the
instrumental resolution widths obtained at the $(3, 1, 4)$ Bragg
peak position. Note the different scale used in the H, K scans and
the L scan. The scans are quite broad in all three directions as
discussed before, and the width does not seem to change very much
below 40 K. In addition, we observe that the maximum peak intensity
occurs not at the lowest temperatures, but at around 15~K. In order
to analyze the temperature dependence in detail, we fitted the H and
K scans to a 2D Lorentzian lineshape, as shown in Fig.~\ref{fig6}.
In Fig.~\ref{fig8}, we present such fitting results for two Q
positions, (3.23, 1, 5.5) and (3, 0.77, 5.5), both around the (3 1
6) Bragg peak. Note that these peaks have CDW modulation vector
$\bf{g}_1$ and $\bf{g}_2$, respectively. Due to the $({\bf Q \cdot
u}_r)^2$ factor, the (3.23, 1, 5.5) peak is much stronger, so that
the intensities plotted in Fig.~\ref{fig8} are scaled to match. On
the other hand, the inverse correlation lengths, $\kappa(T)$, for
these two peaks match exactly without any adjustable parameter. In
Fig.~\ref{fig8}(b), only $\kappa_H$ and $\kappa_K$ of the (3.23, 1,
5.5) peak are shown.

As expected from earlier studies, the CDW peak intensity gradually
decreases and appears to vanish above $\sim 40$~K. In addition to
the peak intensity, we also plotted the integrated intensity in
Fig.~\ref{fig8}(c), which should be proportional to the square of
the CDW order parameter. What is striking in this plot is the
difference between the behavior of the integrated intensity and the
peak intensity. Below about 15 K, the integrated intensity remains
constant, while the peak intensity drops as temperature is
decreased. While the peak intensity almost goes away above 40 K, the
integrated intensity exhibits a cusp around $\sim 42$ K, but
persists up to the structural transition temperature around 55 K. In
order to understand this behavior, one should consider the
temperature dependence of the correlation length. Since integrated
intensity of a Lorentzian peak is given by peak height multiplied by
peak width, the observed decrease in the peak intensity indicates
increase in the peak width. This expectation is borne out by the
fitting results, which shows small increases in the peak widths
below 15 K.

\begin{figure}
\begin{center}
\includegraphics[width=\figwidth]{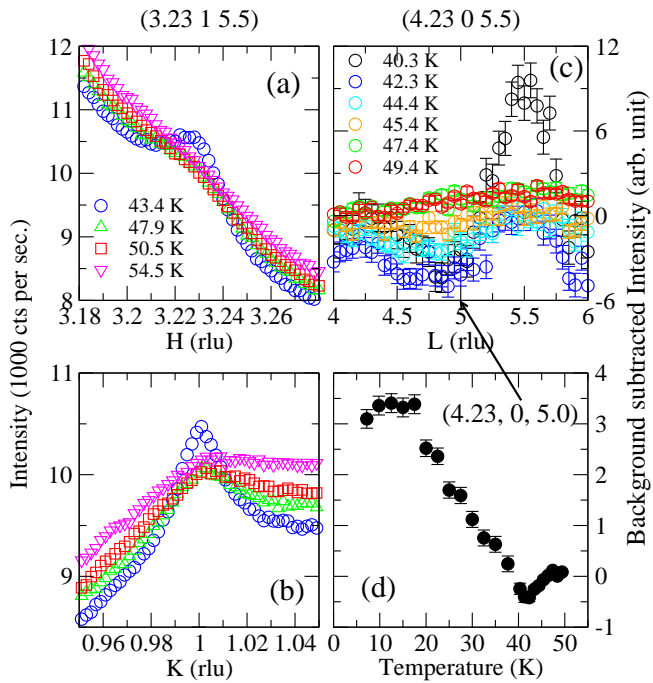}
\end{center}
\caption{(Color online) (a)-(c) H, K, and L scans at high
temperatures, respectively. The H and K scans shown in (a) and (b)
are the same as those shown in Fig.~7, except that these are
obtained at higher temperatures. (c) Background subtracted intensity
around the (4.23, 0, 5.5) peak is shown. The background scan was
taken at 50.4 K, which is justified by the observation that the
L-dependence goes away above $\sim 47$ K. (d) The background
subtracted intensity of (4.23, 0, 5) position as a function of
temperature. This position corresponds to the center point shown in
part (c) and indicates the intensity of the 2D fluctuations.}
\label{fig10}
\end{figure}

\subsection{Short range charge correlation}

Let us first examine the temperature dependence of correlation
lengths above 40 K. In Fig.~\ref{fig10}, scans obtained in this high
temperature range are plotted. The in-plane scans shown in
Fig.~\ref{fig10}(a) and (b) are the same types of scans as the ones
shown in Fig.~\ref{fig7}, but at higher temperatures. One can
clearly see that the peak located at the CDW order wavevector is
still visible at these temperatures, and broadens as temperature
increases. That is, the in-plane correlation length contracts
significantly, and reaches about 16 \AA\ around 55 K, which is
roughly the distance between stripes. Fitting results at these high
temperatures are also plotted in Fig.~\ref{fig8}. In order to model
the temperature dependence of the correlation length, we adopted the
phenomenological expression used in Ref.~\onlinecite{Tranquada99} to
describe spin stripe inverse correlation length:
\begin{equation}
\kappa = \kappa_0 + A e^{-B/k_B T},
\end{equation}
and obtained $\kappa_0=0.0089(2)\AA^{-1}$, $A=35(20)$, and
$B=342(24)$~K. In the second order phase transition, one can fit
$\kappa (T)$ above the transition temperature to a power law,
$\kappa \sim (T-T_c)^\nu$. Due to the limited number of data points
in this regime, our fitting did not yield reliable critical exponent
$\nu$. To compare with Eq. (1), a power law fit with $\nu=1$ is
shown as a dashed line in Fig.~\ref{fig8}(b). Note that the power
law fit seems to describe the data better in the intermediate
temperature range.

In addition, the out-of-plane widths $\kappa_L$ broadens as well,
which is not as noticeable due to the already quite broad L-width.
In order to show this and compare with the in-plane width,
$\kappa_L$ is divided by 8, and plotted as diamonds in
Fig.~\ref{fig8}(b). The out-of-plane width seems to track the
temperature dependence of the in-plane widths, which suggests that
the CDW correlation becomes more two-dimensional as temperature is
raised above $\sim 42$ K. A 2D correlation confined in the ab-plane
will manifests itself as a rod of intensity along the L-direction in
reciprocal space, which can be probed in scattering experiments. In
Fig.~\ref{fig10}(c), the temperature dependence of the L-scans is
shown. In order to follow the temperature dependence of the weak and
broad peak, we plot the background subtracted intensity. There
exists a well-defined peak centered at L=5.5 at T=40.3 K, but this
peak soon falls to the level of background, while the background
intensity increases. As a result, the intensity variation along the
L-direction completely vanishes above $\sim 47$ K, as shown for
$T=47.4$ K and $T=49.4$ K, which means that the CDW correlation at
these temperatures is completely 2D. This type of 2D-3D crossover
can be demonstrated by measuring the temperature dependence of the
background point (L=5), as shown in Fig.~\ref{fig10} (d). Below
$\sim 42$ K, the intensity at this point more or less follows the
peak intensity, since the out-of-plane correlation is quite short
ranged. However, above $\sim 42$ K, this intensity grows as
temperature is increased, saturating above $\sim 47$ K. This
observation seems to indicate that dimensional crossover occurs in
this temperature range between 42~K and 47~K.

In Fig.~\ref{fig8}(c), incommensurability (the peak position) as a
function of temperature is plotted. The incommensurability changes
very little, except for the data point at the highest temperature.
Such change of the incommensurability near the CDW transition is not
very uncommon, and is believed to arise from the short-ranged nature
of the ordering. However, in this case, the change is very small
($\sim 10$ \%), and does not seem to be directly correlated with the
correlation lengths.

\subsection{Low temperature behavior}

Switching our focus to the low temperature range below 40 K, the
peak intensity increases with decreasing temperature and saturates
around 15 K, and actually drops below this temperature. The
correlation length, on the other hand, does not change below $\sim
40$ K. Slight anisotropy in the apparent width along the H and K
direction is due to the instrumental resolution. Since we observe
the same anisotropy ($\kappa_H > \kappa_K$) in both the (3.23, 1,
5.5) peak and the (3, 0.77, 5.5) peak, we conclude that this is
caused by the resolution effect, not by intrinsic anisotropic
correlation of stripes.

As in the case of the high temperature change, the peak intensity
drops below 15 K is largely accounted for by the change in the
correlation length, since the integrated intensity remains more or
less constant. Unfortunately, due to the limitation of the cryostat,
our measurements were limited to temperatures over 8 K. However, we
have repeated the measurements for different CDW peak positions such
as (4.23, 0, 5.5), and also using the Si analyzer, which gives
instrumental H and K resolution of 0.0009 \AA$^{-1}$ and 0.0016
\AA$^{-1}$, respectively. In all our measurements, the observed
behavior could be described consistently by decreasing CDW
correlation length below about 15 K.

\section{Discussion}

According to the temperature dependence study presented in this
work, the following physical picture emerges. Above 55 K, our LBCO
sample is in LTO structure, and no significant charge order
signature is detected. As soon as the LTT structure sets in below 55
K, short range CDW ordered region appears, since the specific
tilting arrangements of the copper oxygen octahedra help stabilizing
the CDW order. These patches of CDW has the same structure as the
fully developed CDW order, extending up to 4 in-plane lattice units,
but there is almost no correlation between next-nearest-neighbor
$\rm CuO_2$ layers. However, given that the in-plane correlation
lengths are isotropic, neighboring layers should satisfy the
crisscrossing pattern. In this sense, this sort of a building block
of stripe order might look like the stripe glass phase recently
reported in Ref.~\onlinecite{Kohsaka07}. However, this high
temperature short-range stripe correlation in LBCO is probably
highly fluctuating. In recent in-plane optical conductivity studies,
a decrease in low energy spectral weight was observed below 60
K.\cite{Homes06} Our observation of the dimensional crossover in CDW
correlation seems to imply that the c-axis optical conductivity may
exhibit different temperature dependence from that of the in-plane
optical conductivity.

As temperature is cooled down further, the correlation lengths in
all three directions expand. In particular, in the temperature range
$T= 42 \sim 47$ K, the 2D CDW correlation crosses over to 3D CDW
order, although the correlation length along the third direction is
very short. The temperature dependence of the correlation length in
this temperature regime can be described with phenomenological
exponential function or a power law. In the latter case, the
critical exponent $0.6 \agt \nu \agt 1.5$ is consistent with our
results. Obviously, current data set is too crude to address the
universality class of the phase transition, and more accurate
measurements are required. However, it is interesting to note that
within error bars, the correlation length behavior is consistent
with random-field Ising model ($\nu \approx 1.5$). We note that in
recent theoretical studies of charge ordering on a square lattice,
the effect of quenched disorder and the resulting random field was
considered.\cite{DelMaestro06} It is interesting to point out that
the short correlation length at high temperature of $\sim 16$ \AA\
roughly corresponds to the average distance between Ba dopants in a
La-O layer.

Below about 40 K, both the peak intensity and the correlation length
grow as temperature decreases until they reach maxima around 15K.
When the sample is cooled further below 15 K, the in-plane
correlation length seems to shrink. In our study of the charge
stripes in applied magnetic field, \cite{LBCO-field} we observed
that the correlation length expands as field is increased above
$\sim 5$ T, indicating that stripe correlation is enhanced as
superconductivity (or fluctuation) is suppressed by either applied
field or elevated temperature. We would like to point out the recent
transport studies by Li et al.,\cite{Li07} in which they observed
anomalous drop in in-plane resistivity around 15 K. It was argued
that two-dimensional superconductivity sets in at this temperature.
Although the relationship between the proposed 2D superconductivity
and suppressed CDW correlation is not clear at the moment, it is
interesting to note that the important temperature (15 K) and field
(5T) scale in both studies match.

In view of recent experimental results, $T \approx 40$ K (perhaps
40-45 K range) appears to be an important temperature scale. Of
course, this temperature is close to the spin density wave ordering
temperature. \cite{Fujita04} In addition, Li et al. observed
anomalous jump in thermopower, resistivity ratio, and magnetic
susceptibility around 40K.\cite{Li07} We add to these our
observation that the CDW correlation crosses over from 3D to 2D, and
the CDW correlation length is saturated below this temperature. This
may suggest that the CDW transition is closely associated with the
SDW transition. In LNSCO, it was observed that the CDW order sets in
at higher temperature than the SDW order, and this observation was
used as evidence that the CDW order drives the SDW order. In LBCO,
on the other hand, it was observed that the CDW and SDW seems to
order at virtually the same temperature.\cite{Fujita04} Our result
suggests that although CDW correlation sets in at higher
temperature, both CDW and SDW develop into a relatively long ranged
order at the same temperature.

One has to be careful when discussing the CDW and SDW ordering
temperatures, since different experimental studies have reported
different transition temperatures of charge ordering in LBCO. In
their neutron scattering study, Fujita and coworkers reported that
the CDW transition and the SDW transition occur at the same
temperature ($\sim 45$ K).\cite{Fujita04} Although the soft x-ray
resonant scattering results show CDW intensity up to 55 K, there
seems to be an indication that the intensity drops around 45 K. We
have measured the temperature dependence utilizing different
instrumental resolution, and observed that the apparent transition
temperature seems to depend on the momentum resolution. Of course,
different samples and different thermometry used in these studies
mean that quantitative comparison is not possible. However, the
overall trend seems to be that the instrumental resolution of the
probing technique does have an effect on the apparent CDW transition
temperature.

One of the important questions arising in the discussion of the
stripe model is whether the charge stripes are site-centered or
bond-centered. This question cannot be answered in our experiments,
since we observe incommensurate CDW order. However, if the
incommensurate superlattice peaks arise due to discommensuration,
the question regarding bond- and site-center remains valid. In this
discommensuration model, one assumes that commensurate charge order
(in this case period of $\lambda_0 =4$) is separated by regularly
spaced domain walls with spacing $d$. If there is a phase slip of
$\phi$ at each domain wall, CDW superlattice peaks do not occur at
the position corresponding to the inverse of the charge order
period: $\epsilon=1/4$. Instead, they are displaced from the
expected superlattice position by the inverse of the large
periodicity including the domain walls: $\epsilon=1/4 + \phi / 2 \pi
d$. In our case, this would mean that there exists a regular array
of domain walls located at about every 12 atoms, or every three
stripes, if we assume $\phi = -\pi/2$. Therefore, such ordered
discommensuration model seems to suggest large number of domain
walls in the CDW ordered region, which may be energetically
unfavorable. Another test of the discommensuration model is the
higher harmonics due to ordered domain walls. Our experimental
results showing that there is no significant higher harmonic
intensity does not support the discommensuration model. Therefore,
the CDW order in our sample seems to be truly incommensurate, which
is consistent with SDW ordering in other 214 materials as discussed
in Ref.~\onlinecite{Robertson06}.

\section{Summary and conclusions}

We have carried out comprehensive x-ray scattering study of charge
stripe ordering in LBCO. We were able to find the following:

\begin{itemize}

\item[1)] We found that the first order structural phase transition from
the so-called LTO phase to the LTT phase occur at 55 K. In this
sample, the LTT phase attains long range order ($\xi > 1400 \AA$) at
low temperatures.

\item[2)] The LTT phase stabilizes stripes along the direction perpendicular
to the tilting of the $\rm CuO_6$ octahedra. We observe only four
satellites around each Bragg peak, which shows that the charge
density wave (CDW) modulation is only compatible with the
one-dimensional stripes.

\item[3)] The CDW ground state is described with one-dimensional sinusoidal
modulation with incommensurate wavevector of (0.23, 0, 0.5). Any
higher harmonics were below our detection limit, which further
supports the incommensurate sinusoidal density wave picture.

\item[4)] We observed that the CDW correlation is isotropic in the $\rm
CuO_2$ plane. That is, the correlation lengths along the
crystallographic a- and b-direction remains the same at all
temperatures. At low temperatures, these are about 200-250 \AA.

\item[5)] The correlation length begins to contract above 40 K in all three
directions. In particular, between 42 K and 47 K, the 3D to 2D
crossover occurs for the CDW correlation. However, short range
isotropic CDW correlation persists up to the LTT-LTO phase
transition temperature.

\item[6)] We were able to explain the L-dependent intensity modulation of
the superlattice peaks by considering displacements of La atoms
along the c-direction.

\end{itemize}

\acknowledgements{We would like to thank J. C. Davis, J. P. Hill, M.
Hucker, E. A. Kim, J. H. Kim, S. Kivelson, M. Lawler, Y. S. Lee, A.
Paramekanti, S. Sachdev, and J. Tranquada for invaluable
discussions. The work at University of Toronto was supported by
Natural Sciences and Engineering Research Council of Canada. The
work at Brookhaven was supported by the U. S. DOE, Office of Science
Contract No. DE-AC02-98CH10886. Use of the Advanced Photon Source
was supported by the U. S. DOE, Office of Science, Office of Basic
Energy Sciences, under Contract No. W-31-109-ENG-38.}

\end{document}